\documentclass[runningheads]{llncs}
\usepackage[T1]{fontenc}
\usepackage{graphicx}
\usepackage{booktabs}
\usepackage[misc]{ifsym}
\newcommand{\corr}{(\Letter)}

\usepackage{amsmath, amssymb}
\usepackage[numbers]{natbib}
\usepackage{float}

\newtheorem{prop}{Proposition}

\tocauthor{Apoorva Lal, Winston Chou}
\toctitle{Estimating Representative Causal Effects with Double Machine Learning}

\begin{document}

\title{Estimating Representative Causal Effects with Double Machine Learning}


\author{Apoorva Lal\inst{1} \and
Winston Chou\inst{2} \corr}

\authorrunning{A. Lal and W. Chou}

\institute{
    OpenAI, San Francisco, CA \email{lal.apoorva@gmail.com}
    \and
    Netflix, New York, NY \email{wchou@netflix.com}
}
\maketitle              

\begin{abstract}
    Double Machine Learning is widely used to estimate treatment effects from non-experimental data.  The ``residuals-on-residuals'' regression (RORR) is especially popular for its simplicity and computational tractability.  However, with heterogeneous treatment effects, the proper interpretation of RORR may not be well understood.  We show that, for non-binary treatments with continuous dose-response functions, RORR estimates a conditional variance-weighted average of derivatives evaluated at treatment values not in the observed dataset.  This estimand does not equal the Average Causal Derivative (ACD) in general.  Hence, even if all units share the same dose-response function, RORR may not estimate an average treatment effect in the population represented by the sample.  We propose an alternative estimator for the ACD that is well suited to the large datasets found in applied data science settings.  We demonstrate the pitfalls of RORR and the favorable properties of the proposed estimator through an illustrative numerical example and with real-world data from Netflix.  Our methodology is used by default in Netflix's observational causal inference platform, where it regularly powers causal research and decision-making at scale.

\end{abstract}

\section{Introduction}

Double Machine Learning (DML) \cite{bickel1993efficient,Chernozhukov2018-fl,robins1994estimation} has become the standard method for estimating causal effects in large, high-dimensional datasets under the assumption of \emph{unconfoundedness}, which postulates that the treatment is as good as randomly assigned given observed covariates \cite{Imbens2004-ir}.  To strengthen the plausibility of unconfoundedness, researchers in many fields use the DML method to condition on numerous covariates when estimating causal effects \cite{chernozhukov2020causal,holtz2020interdependence,jares2025policy}.

DML encompasses a broad class of estimators.  In this paper, we focus on the Partially Linear Model (PLM), which relates an outcome $Y_i$ to a continuous or discrete-valued treatment $T_i$ conditional on pretreatment covariates $X_i$ by
\begin{equation}
    Y_i = \theta T_i + g(X_i) + e_i \quad \text{and} \quad T_i = h(X_i) + u_i.
\end{equation}
The PLM imposes minimal assumptions on how the treatment and outcome relate to covariates.  It also motivates an intuitive two-step DML estimator of $\theta$, the residuals-on-residuals regression (RORR).  RORR, an extension of the Frisch-Waugh-Lovell theorem, first ``partials out'' the effect of $X_i$ using flexible machine learning methods; then forms the residuals
\begin{equation}
\widetilde{Y}_i = Y_i - \widehat{m}(X_i) \quad \text{and} \quad \widetilde{T}_i = T_i - \widehat{h}(X_i),
\end{equation}
where $\widehat{m}$ and $\widehat{h}$ are estimators for $m(X_i) := E[Y_i | X_i]$ and $h(X_i) := E[T_i | X_i]$, respectively; and lastly regresses $\widetilde{Y}_i$ on $\widetilde{T}_i$ to obtain the RORR estimate of $\theta$, which we denote by $\hat{\theta}$ \cite{Robinson1988-dm}. 

When the treatment effect $\theta$ is the same for all units in the population, it is also the Average Treatment Effect (ATE) for binary treatments, the Average Causal Derivative (ACD) for continuous treatments, and the Average Incremental Effect (AIE) for integer-valued treatments.  Alternatively, we study the probability limit (plim) and interpretation of $\hat{\theta}$ when treatment effects are \emph{heterogeneous}.  Under such heterogeneity, $\hat{\theta}$ converges to a conditional variance-weighted average of causal effects, which assigns greater weight to units whose treatment values are less predictable.  For example, when $T_i$ is binary, the RORR estimand assigns the greatest weight to units whose conditional probability of treatment is closest to 0.5.  This biases it away from the ATE in general \cite{Angrist1998-ok, Aronow2016-nn, Sloczynski2022-kg}.  When treatments are non-binary (for example, continuous), RORR is subject to the more nuanced biases explored in this paper.

We contribute a general analysis of RORR and its biases with both binary and non-binary treatments.  We demonstrate the empirical relevance of these biases with both a stylized numerical example and real-world data from Netflix.  We propose a simple alternative estimator based on Augmented Inverse Propensity Weighting (AIPW), which coarsens the treatment into bins \cite{Cattaneo2010-oc, kennedy2017non}.  We establish the favorable theoretical properties of the estimator and apply it to empirical data.  Our coarsened AIPW estimator has been used for observational causal inference and decision-making at Netflix for several years, and is the default methodology deployed on Netflix's observational causal inference platform.

\section{Residuals-on-Residuals Regression with Heterogeneous Treatment Effects}

The residuals-on-residuals regression (RORR) is a popular observational causal inference tool for many reasons.  Most importantly, it enables the use of highly flexible modern machine learning estimators for the nuisances in the PLM, thus freeing researchers from making strong assumptions about those parameters.  If the PLM is correctly specified, it is statistically efficient and can achieve parametric convergence rates even when the nuisances are estimated nonparametrically.  Attesting to its popularity, recent applications of RORR can be found in economics \cite{BaiardiNaghi2024}, ecology \cite{FinkEtAl2023}, political science \cite{jares2025policy}, and public health \cite{WeiEtAl2024}.

In applied data science settings---characterized by large datasets, short timelines, and stakeholders with diverse technical backgrounds---RORR is also commonly used for practical reasons.  For example, the final regression step is computationally efficient, as only a few statistics are needed to compute $\hat{\theta}$.  Furthermore, the two-step recipe of (1) removing variation explained by pretreatment covariates and (2) estimating the effect of the remaining exogenous variation in $T_i$ on $Y_i$ is intuitive and easy to explain to non-experts.  In its simplest form, RORR is none other than ordinary least squares \cite{robins2007comment}.

However, this simplicity comes at a price.  The interpretability of $\hat{\theta}$ as an estimate of the ``average'' treatment effect (whether the ATE, ACD, or AIE) depends on the assumption of a homogeneous treatment effect in the PLM, which may not hold in applications.  Below, we discuss the interpretation of $\hat{\theta}$ under two violations of this assumption: binary treatments with heterogeneous causal effects across individual units (e.g., a price discount whose impact is greater for more price-sensitive customers), and non-binary treatments with nonlinear dose-response functions (e.g., the effect may depend on the amount that is discounted).

\subsection{Binary Treatments with Heterogeneous Treatment Effects}

We begin with the RORR estimand for binary treatments with heterogeneous treatment effects.  Letting $T_i \in \{0, 1\}$ denote the binary treatment, we consider the model
\[
    Y_i = \theta_i T_i + g(X_i) + e_i \quad \text{and} \quad T_i = h(X_i) + u_i,
\]
where $\theta_i$ is an individual treatment effect.  We assume unconfoundedness of the treatment given $X_i$, which in turn implies that the errors are conditionally exogenous and uncorrelated: $E[e_i|X_i] = 0$, $E[u_i|X_i] = 0$, and $E[e_i u_i|X_i] = 0$.  Thus, $g(X_i) = E[Y_i - \theta_i T_i | X_i]$ and $h(X_i) = E[T_i | X_i]$.  Unconfoundedness further implies that $\theta_i$ is conditionally independent of $T_i$ given $X_i$.  Lastly, we assume that the treatment is not deterministically assigned: $E[(T_i - h(X_i))^2] > 0$.

We analyze the plim of the OLS regression of $Y_i - m(X_i)$ on $T_i - h(X_i)$ with i.i.d. observations $(Y_i, T_i, X_i)$, $i = 1, \ldots N$.  Although $m$ and $h$ must be estimated in practice, we focus on the (true) limiting $m$ and $h$ to emphasize that the RORR plim is biased for the ATE even when the researcher has access to sufficient covariates and consistently estimates these nuisances.\footnote{In practice, researchers often use extremely flexible function classes for $m$ and $h$ (e.g., deep neural networks) that are able to approximate the true nuisance functions arbitrarily closely.}

First, observe that
\begin{eqnarray}
    \hat{\theta} \overset{p}{\to} \frac{E[(T_i - h(X_i)) (Y_i - m(X_i))]}{E[(T_i - h(X_i))^2]} = \frac{E[\theta_i (T_i^2 - T_i h(X_i))]}{E[(T_i - h(X_i))^2]}.
\end{eqnarray}
Using the fact that $T_i$ is binary and applying the law of iterated expectations, we rewrite this as
\begin{eqnarray}
    \frac{E[\theta_i (T_i^2 - T_i h(X_i))]}{E[(T_i - h(X_i))^2]} &=& \frac{E[E[\theta_i | X_i] E[T_i^2 - T_i h(X_i) | X_i]]}{E[(T_i - h(X_i))^2]} \label{eqn:theta_indep} \\ \nonumber
    &=& \frac{E[\theta_i (T_i - h(X_i))^2]}{E[(T_i - h(X_i))^2]},
\end{eqnarray}
where Equation~(\ref{eqn:theta_indep}) follows from $\theta_i$ being conditionally independent of $T_i$.  This establishes the well-known result that, with a binary treatment, linear regression converges to a conditional variance-weighted average of individual treatment effects \cite{Angrist1998-ok,Aronow2016-nn,Sloczynski2022-kg}.

For an intuitive restatement of this result, denote the conditional variance weights by $\omega_i := \frac{(T_i - h(X_i))^2}{E[(T_i - h(X_i))^2]}$ and note that $E[\omega_i] = 1$ by construction.  Then the bias of the RORR plim (which we will denote by $\tilde{\theta}$) with respect to the ATE can be written as
\begin{eqnarray}
    \tilde{\theta} - E[\theta_i] = E[\omega_i \theta_i] - E[\omega_i] E[\theta_i] = \operatorname{Cov}(\omega_i, \theta_i).
    \label{eqn:binary-bias}
\end{eqnarray}
In other words, the RORR bias for binary treatments when treatment effects are heterogeneous is proportional to the covariance of the individual treatment effects with the residual treatment variance.  This covariance need not equal zero, and therefore $\tilde{\theta} \neq E[\theta_i]$ in general.\footnote{A corollary is that ranking treatments based on their PLM coefficient is not the same as ranking them based on their ATEs \cite{lal2024doesregressionproducerepresentative}.}

\subsection{Non-Binary Treatments}

\label{sec:many}

We now turn our attention to non-binary, many-valued (e.g., continuous or integer-valued) treatments with nonlinear dose-response functions.  Although past research has examined the effect of treatment effect heterogeneity on the interpretation of linear treatment effect estimators, it has mainly done so in the context of binary treatments and/or linear treatment effects \cite[][]{Aronow2016-nn}.  However, in many applications, treatments are many-valued and/or have nonlinear effects on the outcome---for example, they may have diminishing returns.  Such nonlinearity is an empirically common form of treatment effect heterogeneity \cite[][]{angrist1999empirical, Yitzhaki1996-io}.  Here, we present a bias decomposition for RORR with non-binary treatments that, to our knowledge, is novel in the literature and illuminates its similarities with and differences from the binary case.

Specifically, we study the model
\begin{equation}
    Y_i = f(T_i) + g(X_i) + e_i \quad \text{and} \quad T_i = h(X_i) + u_i, \label{eqn:continuous-plm}
\end{equation}
where $f$ is a twice-continuously differentiable function of some non-binary treatment $T_i$.  As before, we assume conditional ignorability, consistent estimators for $m$ and $h$, non-deterministic treatment, and i.i.d. observations.

Under these assumptions, the RORR estimate converges to
\begin{eqnarray}
    \hat{\theta} &\overset{p}{\to}& 
    \frac{E[(T_i - h(X_i)) f(T_i)]}{E[(T_i - h(X_i))^2]}.
\end{eqnarray}
Since $h(X_i)$ is a constant given $X_i$ and applying the law of iterated expectations, we can rewrite this as
\begin{eqnarray}
    \frac{E[E[(T_i - h(X_i)) (f(T_i) - f(h(X_i))) | X_i]]}{E[(T_i - h(X_i))^2]}.
\end{eqnarray}
By the mean value theorem, there exists a $T_i^*$ between $T_i$ and $h(X_i)$ for all $X_i$ such that
\begin{eqnarray}
    \label{eqn:many-rorr-plim}
    \frac{E[E[(T_i - h(X_i)) (f(T_i) - f(h(X_i))) | X_i]]}{E[(T_i - h(X_i))^2]} &=& \frac{E[(T_i - h(X_i))^2 f'(T_i^*)]}{E[(T_i - h(X_i))^2]} \label{eqn:non-binary-bias} \\ \nonumber
     &=& \qquad E[\omega_i f'(T_i^*)].
\end{eqnarray}
\eqref{eqn:non-binary-bias} shows that, as in the binary treatment setting, $\hat{\theta}$ also converges to a conditional variance-weighted average of causal effects.\footnote{Not coincidentally, the Wald estimand with a binary instrument and a continuous endogenous treatment can also be represented as a weighted average of derivatives at the mean values $T_i^*$ \cite{angrist1999empirical}.}  However, unlike in the binary treatment case, the quantity being averaged cannot be interpreted as the causal effect of increasing the treatment in the population represented by the sample.  This is because the mean value $T_i^*$ is not the actual treatment dose received by $i$, but a convex combination of the realized treatment $T_i$ and its conditional mean. As such, $T_i^*$ may not be an observed treatment level.  If $T_i$ is not continuous, it may not even be a realizable treatment value. 

Proposition~\ref{prop:many} establishes the restrictive conditions under which $\tilde{\theta}$ converges to the ACD.
\begin{prop}
    \label{prop:many}  Let $(Y_i, T_i, X_i)$ be i.i.d. draws from a distribution obeying the structural model~(\ref{eqn:continuous-plm}).  Assume conditional ignorability, $f$ twice-continuously differentiable on its domain, $f'$ Lipschitz with minimal constant $L$, non-deterministic treatment, $E[|T_i - h(X_i)|^3]$ finite, and consistent estimators of $m$ and $h$.  Then the RORR plim
    $$
    \tilde{\theta} = \frac{E[(T_i - h(X_i)) f(T_i)]}{E[(T_i - h(X_i))^2]}
    $$
    equals the Average Causal Derivative $E[f'(T_i)]$ if $f$ is affine ($L = 0$).  More generally, its bias decomposes into a curvature term and a conditional-variance weighting term, which need not vanish unless $f$ is affine.
\end{prop}

\begin{proof}
With i.i.d. observations, conditional ignorability, and $m$ and $h$ consistently estimated, $\tilde{\theta}$ is as given in~(\ref{eqn:many-rorr-plim}).  We decompose the bias of $\tilde{\theta}$ relative to the ACD into
\begin{eqnarray}
    \label{eqn:bias-decomposition}
    && E[\omega_i f'(T_i^*)] - E[f'(T_i)] \\ \nonumber
    && \qquad = \underbrace{E[\omega_i f'(T_i^*)] - E[\omega_i f'(T_i)]}_{:= A}
    + \underbrace{E[\omega_i f'(T_i)] - E[f'(T_i)]}_{:= B}.
\end{eqnarray}
$A$ is the difference between the RORR plim and the weighted average causal derivative evaluated over the treatment distribution actually observed in the sample.  Rewrite $A$ as
\begin{eqnarray}
    \frac{E[(T_i - h(X_i))^2 (f'(T_i^*) - f'(T_i))]}{E[(T_i - h(X_i))^2]}.
\end{eqnarray}
Using the Lipschitz property of $f'$,
\begin{equation}
    |f'(T_i^*) - f'(T_i)| \le L|T_i^* - T_i| \le L|T_i - h(X_i)|,
\end{equation}
where the last inequality follows from $T_i^*$ lying between $T_i$ and $h(X_i)$.  Multiplying both sides by $(T_i - h(X_i))^2$ and taking expectations, we have
\begin{eqnarray}
 E[(T_i - h(X_i))^2 |f'(T_i^*) - f'(T_i)|] \le L E[|T_i - h(X_i)|^3].
\end{eqnarray}
Dividing both sides by $E[(T_i - h(X_i))^2]$ yields the bound
\begin{equation}
    |A| \le L \underbrace{\frac{E[|T_i - h(X_i)|^3]}{E[(T_i - h(X_i))^2]}}_{:= \kappa}. \label{eqn:bound}
\end{equation}
Non-deterministic treatment implies that $\kappa$ exists and is strictly positive.  Therefore, the bound on $|A|$ is 0 if $f$ is affine ($L = 0$).

The second bias term $B$ has the same interpretation as in the binary treatment case.  That is, letting $\omega_i$ again denote the conditional variance weight,
\begin{equation}
    E[\omega_i f'(T_i)] - E[f'(T_i)] = \operatorname{Cov}(\omega_i, f'(T_i)),
\end{equation}
which is the continuous analogue of Equation~(\ref{eqn:binary-bias}).  If $f$ is affine, $f'(T_i)$ is a constant, so $\operatorname{Cov}(\omega_i, f'(T_i)) = 0$. Therefore, the absolute bias of $\tilde{\theta}$ is bounded by
\begin{eqnarray}
    |A + \operatorname{Cov}(\omega_i, f'(T_i))| &\le& |A| + |\operatorname{Cov}(\omega_i, f'(T_i))| \\ \nonumber
    &\le& L\kappa + |\operatorname{Cov}(\omega_i, f'(T_i))|, 
\end{eqnarray}
which equals 0 if $f$ is affine.  This obtains the result. \hfill \qed
\end{proof}

In other words, the bias of $\tilde{\theta}$ for the ACD decomposes into two terms.  The first depends on the curvature of $f$, which ``pulls'' the derivatives being averaged away from the derivatives at the realized treatment values.  The second term reflects conditional variance-weighting and has the same interpretation as in the binary treatment case.  The first term is eliminated when $f$ is affine.  The second term is eliminated if there is no treatment heterogeneity, which holds trivially when $f$ is affine.  Therefore, $A$ and $B$ vanish when $f$ is affine (and thus the PLM is correctly specified).  However, if $f$ is not affine, then the biases need not vanish except in contrived cases (e.g., they cancel exactly).

\section{Numerical Example}
\label{sec:numerical}

To build intuition, we walk through a stylized numerical example.\footnote{Replication code for all figures and tables in this section can be found at \url{https://github.com/winston-chou/rorr}.} Although we make simplifying assumptions to enable closed-form analysis, our choices also reflect common aspects of real-world data.

Specifically, we assume:
\begin{enumerate}
    \item \emph{Diminishing returns.} While $E[Y_i | T_i, X_i]$ is increasing in $T_i$, it also exhibits diminishing returns.  That is, $f'(T_i) > 0$ and $f''(T_i) < 0$.
    \item \emph{Right-skewed treatments.} $T_i$ is an overdispersed count variable, such that even a correct model for $E[T_i | X_i]$ has heteroskedastic errors.
\end{enumerate}

Let $X_i$ be a categorical variable that takes on values $j = 1, \ldots, J$ with probabilities $\pi_1, \ldots, \pi_J$.  Let $T_i$ be conditionally Poisson given $X_i$ with parameters $\lambda_1, \ldots, \lambda_J$.  Let $f(T_i) = \log(T_i + 1)$.  Then the conditional expectation of the causal derivative of $Y_i$ with respect to $T_i$ is
\begin{equation}
 E[f'(T_i) | X_i = j] = \frac{1 - \exp(-\lambda_j)}{\lambda_j}
 \label{eqn:poisson-drv}
\end{equation}
and the ACD is $\sum_j \pi_j \frac{1 - \exp(-\lambda_j)}{\lambda_j}$.  (See Appendix~A of the supplementary material for derivations.)

The RORR plim is
\begin{equation}
 \hat{\theta} \overset{p}{\to} \tilde{\theta} = \frac{\sum_j^J \pi_j E[(T_i - \lambda_j)^2 f'(T_i^*) | X_i =j]}{\sum_j^J \pi_j \lambda_j},
 \label{eqn:rorr}
\end{equation}
where, as before, $T_i^*$ is a point between $T_i$ and $\lambda_j$.  Note the two biases relative to the ACD.  First, rather than evaluate $f'$ at $T_i$, we evaluate it at $T_i^*$.  Second, we also weight each $f'(T_i^*)$ by its normalized conditional variance,
\begin{equation}
\omega_i = (T_i - \lambda_{j[i]})^2 / \sum_j \pi_j \lambda_{j[i]}.
\end{equation}

\begin{figure}
    \centering
    \includegraphics[width=2.5in, keepaspectratio]{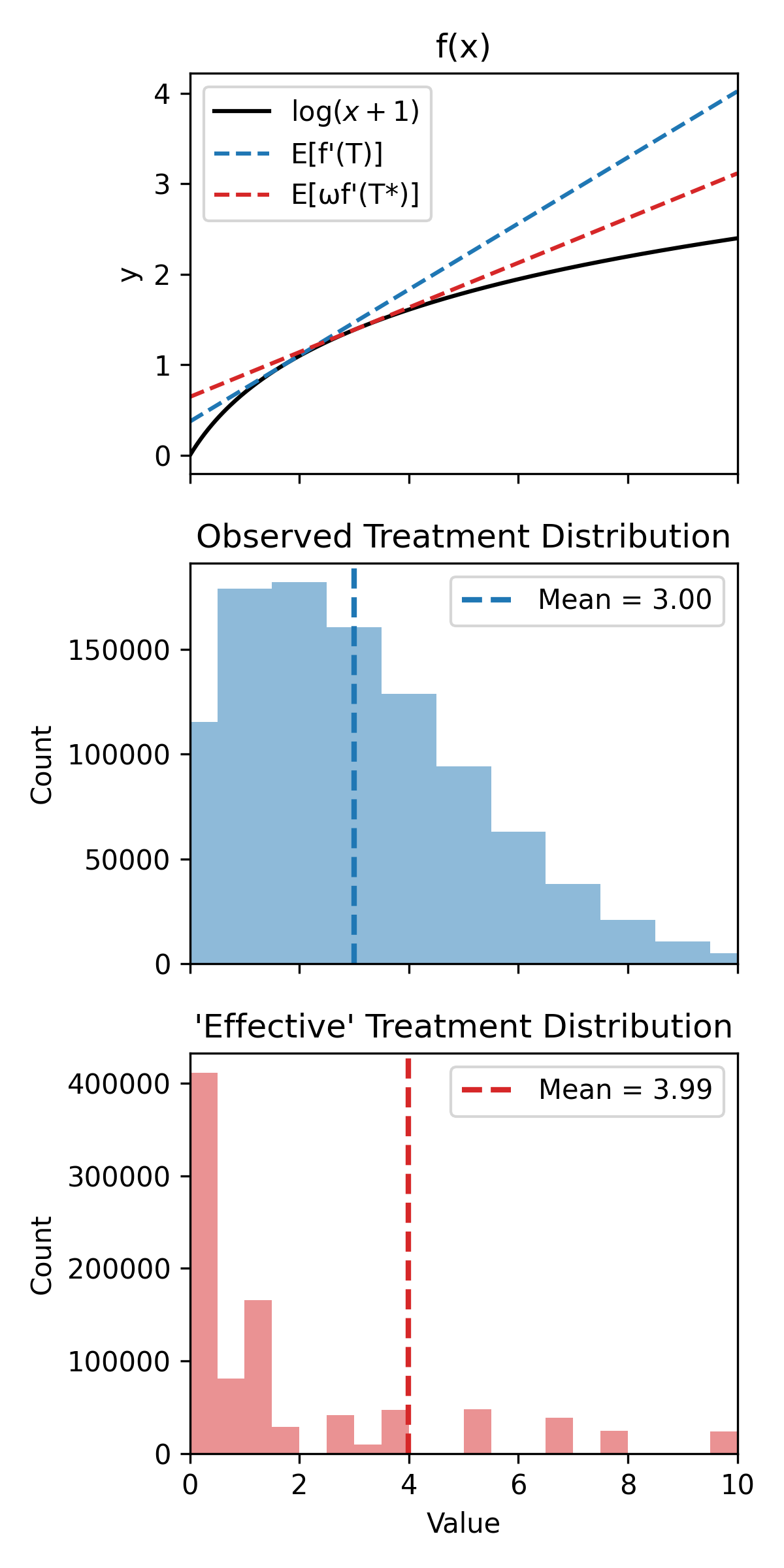}
    \caption{Bias of Residuals-on-Residuals Regression (RORR). RORR undershoots the true ACD due to the concavity of the dose-response function and the rightward shift of the ``effective'' treatment distribution.}
    \label{fig:bias}
\end{figure}

Figure~\ref{fig:bias} illustrates the resulting bias.  In the top panel, we plot the dose-response curve $f(T_i) = \log(T_i + 1)$, as well as tangent lines with slopes equal to $E[f'(T_i)]$ in blue and to $E[\omega_i f'(T_i^*)]$ in red, where $T_i^*$ is the ``effective'' treatment analyzed by RORR.  The key takeaway is that RORR targets a quantity different from, and smaller than, the ACD.\footnote{An analogous result in the welfare economics literature is that OLS upweights the slopes of higher-income groups in regressions of consumption on income, leading to attenuation bias \cite{Yitzhaki1996-io}.}  The subsequent panels give intuition for this result.  After weighting by $\omega_i$ and transforming $T_i$ to $T_i^*$, the effective treatment distribution is much more right-skewed than the observed treatment distribution.  This means that we tend to evaluate the slope of $f$ at higher values of $T_i$.  This induces negative bias because $f''(T_i) < 0$.

In Table~\ref{tab:rorr_acd}, we report the empirical RORR estimated in simulations at varying sample sizes.  For comparison, we also report the empirical ACD (calculated as the sample mean of $\frac{1}{T_i + 1}$) and the true ACD computed using~(\ref{eqn:poisson-drv}).\footnote{Note that, because $T_i$ is integer-valued in this example, the more appropriate causal estimand is the Average Incremental Effect (AIE), defined as
\begin{equation}
    E[Y_i(T_i + 1) - Y_i(T_i)] = \sum_{t=0}^\infty [f(t + 1) - f(t)] \Pr(T_i = t).
\end{equation}
However, because the RORR plim is a weighted average of derivatives, we focus on the ACD in Table~\ref{tab:rorr_acd} for comparability and defer discussion of the AIE to Section~\ref{sec:aipw}.}  The results in the table confirm the negative bias of RORR for the ACD.

\begin{table}
\centering
\setlength{\tabcolsep}{10pt}
\begin{tabular}{lrrr}
\toprule
Sample Size & Empirical RORR & RORR 95\% CI & RORR Plim \\
\midrule
10,000 & 0.250 & (0.237, 0.262) & 0.247 \\
100,000 & 0.248 & (0.244, 0.252) & 0.247 \\
1,000,000 & 0.248 & (0.247, 0.249) & 0.247 \\
\midrule
Sample Size & Empirical ACD & ACD 95\% CI & True ACD \\
\midrule
10,000 & 0.362 & (0.357, 0.367) & 0.365 \\
100,000 & 0.364 & (0.362, 0.366) & 0.365 \\
1,000,000 & 0.365 & (0.364, 0.365) & 0.365 \\
\bottomrule
\end{tabular}
\medskip
\caption{Simulation Results for RORR and Average Causal Derivative (ACD).}
\label{tab:rorr_acd}
\end{table}

\paragraph{Practical guidance.} Given the prevalence of diminishing returns and right-skewed treatments in applied work, our theoretical analysis suggests that, as a rule of thumb, the RORR estimate will be negatively biased for the ACD even when the unconfoundedness assumption is satisfied.  Again, this is because the combination of right-skewedness and diminishing returns means that the units that RORR implicitly up-weights will tend to have higher values of the treatment, where the dose-response curve is flatter.  We propose a consistent estimator for the ACD in Section~\ref{sec:aipw} and give real-world evidence for this rule of thumb in Section~\ref{sec:empirical}.

\section{Coarsened Augmented Inverse Propensity Weighting}

\label{sec:aipw}

A benchmark estimator for the Average Causal Derivative (ACD) with continuous treatments uses the Generalized Propensity Score (GPS) \cite{Imbens2000-sk}. However, GPS estimators require estimating the conditional density of the treatment, which can suffer from slow rates and instability \cite{kennedy2017non}.

As an alternative to both RORR and GPS, we propose a simple coarsened Augmented Inverse Propensity Weighting (AIPW) estimator, which uses the AIPW estimates of counterfactual means as building blocks \cite[cf.][]{Cattaneo2010-oc, kennedy2017non}.  This estimator proceeds by first partitioning the treatment support into $K$ disjoint bins of length $\ell$.  Then, for each bin $S_k$, we estimate the bin-level propensity score $p_k(x) = \Pr(T_i \in S_k | X_i = x)$. Denote this estimate by $\hat{p}_k$.  For each treatment bin, we also estimate a conditional mean regression denoted by $\hat{m}_k(X_i)$, which estimates $m_k(x) = E[Y_i | T_i \in S_k, X_i = x]$.

Next, we form the usual AIPW estimator for the marginal counterfactual mean in bin $S_k$
\begin{equation}
\hat{\psi}_k := \frac{1}{N}\sum_{i=1}^N \left(\left[\frac{\mathbf{1}(T_i \in S_k)}{\hat{p}_k(X_i)}(Y_i - \hat{m}_k(X_i))\right] + \hat{m}_k(X_i)\right).
\end{equation}
This estimator is doubly-robust, meaning that if either $\hat{p}_k$ or $\hat{m}_k$ is consistent, then $\hat{\psi}_k$ is also consistent for the mean potential outcome within $S_k$ \citep{kennedy2024semiparametric}.\footnote{The plim of $\hat{\psi}_k$ has a nuanced interpretation due to averaging over the bin.  Under strong ignorability, it can be interpreted as the population-average potential outcome if units are treated according to the conditional distribution of $T_i$ given $X_i$ and $T_i \in S_k$.}

Then the \emph{coarsened AIPW} estimate of the ACD is given by
\begin{equation}
\hat{\psi} := \sum_{k=1}^{K-1} w_k \left(\frac{\hat{\psi}_{k + 1} - \hat{\psi}_k}{\overline{t}_{k + 1} - \overline{t}_k}\right),
\label{eqn:coarsened-aipw}
\end{equation}
where $\overline{t}_k = \frac{t_{k+1} + t_k}{2}$, $\pi_k = \Pr(T_i\in S_k)$, and $w_k=\frac{\pi_k}{\sum_{r=1}^{K-1}\pi_r}$. Heuristically, $\hat{\psi}$ approximates $f'$ using a piecewise linear function and computes its weighted average using the empirical distribution of the lower segment. 

Theorem~\ref{thrm:aipw} proves the consistency of this estimator under regularity conditions.  For the proof, we introduce the potential outcomes notation
\begin{equation}
    Y_i(t) = f(t) + g(X_i) + e_i \quad \text{and} \quad T_i = h(X_i) + u_i.
    \label{eqn:potential-outcomes}
\end{equation}

\begin{theorem}
\label{thrm:aipw}
Let $(Y_i, T_i, X_i)$ be i.i.d. draws from a distribution that obeys the structural model~\eqref{eqn:potential-outcomes}. Assume that $T_i$ is distributed on a compact interval $\mathcal T=[\underline t,\overline t]$ and define the coarsened AIPW estimator $\widehat{\psi}$ as in~\eqref{eqn:coarsened-aipw}.

Assume unconfoundedness and the following positivity condition:
\begin{equation}
p_k(x)\ge \epsilon \ell,
\end{equation}
for all $k$ and $x$, for some $\epsilon > 0$.  Assume that the conditional treatment density $p(t\mid x)$ is twice continuously differentiable and has uniformly bounded derivatives.  As before, assume that $f'(t)$ is Lipschitz continuous.  Lastly, assume that the nuisances are estimated (e.g., with cross-fitting) such that the bin-level AIPW estimates are uniformly consistent in $k$
\begin{equation}
\max_{1\le k\le K}|\widehat{\psi}_k-\psi_k|=o_p(1).
\end{equation}

Then, as $N\to \infty$, $K\to \infty$, and $\ell\to 0$,
\begin{equation}
\widehat{\psi} \xrightarrow{p} E[f'(T_i)].
\end{equation}
\end{theorem}

\begin{proof}
We begin by showing that $\widehat{\psi} - \psi = o_p(1)$, where
\begin{equation}
\widehat{\psi}
=
\sum_{k=1}^{K-1}
w_k
\frac{\widehat{\psi}_{k+1}-\widehat{\psi}_k}{\ell}
\qquad \text{and} \qquad
\psi
=
\sum_{k=1}^{K-1}
w_k
\frac{\psi_{k+1}-\psi_k}{\ell}.
\end{equation}
First, write
\begin{equation}
\widehat{\psi} - \psi
=
\sum_{k=1}^{K-1}
w_k
\frac{\varepsilon_{k+1}-\varepsilon_k}{\ell},
\end{equation}
where $\varepsilon_k=\widehat{\psi}_k-\psi_k$.  Note that $\varepsilon_k = o_p(1)$ by assumption.  Rearranging terms, we have
\begin{equation}
\widehat{\psi} -\psi
=
-\frac{w_1}{\ell}\varepsilon_1
+
\sum_{k=2}^{K-1}
\frac{w_{k-1}-w_k}{\ell}\varepsilon_k
+
\frac{w_{K-1}}{\ell}\varepsilon_K.
\end{equation}

Boundedness and continuous differentiability of $p(t\mid x)$ imply that the marginal treatment density $\pi(t)$ is also bounded and continuously differentiable (and, in particular, that $\pi_k$ is bounded and continuously differentiable uniformly in $k$).  This in turn implies that $w_k = \frac{\pi_k}{1 - \pi_K} = O(\ell)$, so $\frac{w_1}{\ell}$ and $\frac{w_{K-1}}{\ell}$ are both $O(1)$.  Continuous differentiability of $\pi(t)$ further implies that $|w_k - w_{k - 1}| = O(\ell^2)$ uniformly in $k$, so
\begin{equation}
\sum_{k=2}^{K-1}\left|\frac{w_{k-1}-w_k}{\ell}\right| = O(\ell^{-1}) \cdot O(\ell) =O(1).
\end{equation}
Therefore,
\begin{equation}
|\widehat{\psi} - \psi|
\le
O(1) \cdot \max_{1\le k\le K}|\varepsilon_k|
=
o_p(1).
\label{eqn:estimation}
\end{equation}

Next, we consider the error of the coarsened population estimand $\psi$ for $E[f'(T_i)]$. Define
\begin{equation}
r_k(t\mid x)
=
\frac{p(t\mid x)1(t\in S_k)}{p_k(x)}.
\end{equation}
Under unconfoundedness,
\begin{equation}
m_k(x)
=
\int_{S_k}
\{f(t)+g(x)\}r_k(t\mid x)\,dt,
\end{equation}
and so
\begin{equation}
\psi_k
=
E[g(X_i)]
+
E\left[
\int_{S_k} f(t)r_k(t\mid X_i)\,dt
\right].
\end{equation}
Define
\begin{eqnarray}
    \Delta_k := \psi_{k+1} - \psi_k = \int \bigg(\int f(t) r_{k+1}(t|x) dt - \int f(t) r_k(t|x) dt\bigg) p(x) dx.
\end{eqnarray}

Lemma 1 in Appendix C of the supplementary material establishes that the inner integrals
\begin{equation}
\int f(t) r_k(t|x) dt = f(\overline{t}_k) + O(\ell^2),
\end{equation}
uniformly in $x$, so after integrating over $p(x)$, we obtain
\begin{equation}
\Delta_k = f(\overline{t}_{k + 1}) - f(\overline{t}_k) + O(\ell^2).
\end{equation}
By the mean value theorem, there exists a $\xi_k \in [\overline{t}_k, \overline{t}_{k+1}]$ such that
\begin{eqnarray}
    \Delta_k = f'(\xi_k) \underbrace{(\overline{t}_{k+1} - \overline{t}_k)}_{=\ell} + O(\ell^2),
\end{eqnarray}
so $\Delta_k / \ell = f'(\xi_k) + O(\ell)$.  Thus, we have that
\begin{equation}
    \psi = \sum_{k=1}^{K-1} w_k\frac{\Delta_k}{\ell} = \sum_{k=1}^{K-1} w_k f'(\xi_k) + O(\ell).
    \label{eqn:riemann}
\end{equation}

It remains to show that the weighted sum in~\eqref{eqn:riemann} approximates $E[f'(T_i)]$. By definition of $w_k$,
\begin{equation}
\sum_{k=1}^{K-1}w_k f'(\xi_k)
=
\frac{1}{1-\pi_K}
\sum_{k=1}^{K-1}\pi_k f'(\xi_k).
\end{equation}
Because $f'$ is Lipschitz and $\xi_k$ lies within $O(\ell)$ of every point
in $S_k$,
\begin{equation}
\sum_{k=1}^{K-1}\pi_k f'(\xi_k)
=
\sum_{k=1}^{K-1}
\int_{S_k} f'(t)p(t)\,dt
+
O(\ell).
\end{equation}
Because $\pi_K$ is $O(\ell)$, $(1 - \pi_K)^{-1}$ is $1 + O(\ell)$, which does not change the order of the approximation.  Therefore,
\begin{equation}
\sum_{k=1}^{K-1}w_k f'(\xi_k)
=
E[f'(T_i)]
+
O(\ell).
\label{eqn:discretization}
\end{equation}
Combining \eqref{eqn:estimation} and \eqref{eqn:discretization} gives
\begin{equation}
\widehat{\psi} = E[f'(T_i)]
+
O(\ell)
+
o_p(1),
\end{equation}
which converges in probability to $E[f'(T_i)]$ as $\ell\to 0$. \hfill \qed
\end{proof}

$\Delta_k$ has a subtle interpretation: it is the ``effect'' of moving from segment $S_k$ to $S_{k+1}$ when units are treated probabilistically according to the conditional treatment distribution in each segment.  It can also be interpreted as the ATE of a deterministic intervention. Given continuity of $f$, $p$, and positivity, the mean value theorem for integrals states that there exists a $\tilde{t}_k \in S_k$ such that
\begin{equation}
    \int f(t) \frac{p(t | x) 1(t \in S_k)}{\Pr(t \in S_k | x)} dt = f(\tilde{t}_k).
    \label{eqn:mvt-integral}
\end{equation}
Note that this $\tilde{t}_k$ implicitly depends on $x$, which we have suppressed from our notation for readability.

Thus, we can rewrite \eqref{eqn:mvt-integral} as
\begin{eqnarray}
    \Delta_k = \int [f(\tilde{t}_{k+1}) - f(\tilde{t}_k)] p(x) dx,
\end{eqnarray}
showing that $\Delta_k$ is the ATE of fixing $T_i$ to some covariate-dependent $\tilde{t}_{k+1} \in S_{k+1}$ relative to $\tilde{t}_k \in S_k$.

Table~\ref{tab:results} shows the results of applying this estimator to the simulated data from Section~\ref{sec:numerical}.  Because the treatment is integer-valued, we can set the bins to each observed treatment value.  As the table shows, this yields a consistent estimator for the AIE.\footnote{Note that the AIE is less than the ACD because $f(t + 1) - f(t) = \log\left(1 + \frac{1}{t+ 1}\right) \le \frac{1}{t + 1}$ for all $t \ge 0$.}

\begin{table}
\centering
\setlength{\tabcolsep}{10pt}
\begin{tabular}{lrrr}
\toprule
Sample Size & Empirical AIE & AIE 95\% CI & True AIE \\
\midrule
10,000 & 0.277 & (0.247, 0.308) & 0.295 \\
100,000 & 0.291 & (0.282, 0.300) & 0.295 \\
1,000,000 & 0.295 & (0.290, 0.300) & 0.295 \\
\bottomrule
\end{tabular}
\medskip
\caption{Consistency of AIPW for the Average Incremental Effect (AIE).}
\label{tab:results}
\end{table}

\paragraph{Practical guidance.}  Our proof of the consistency of the coarsened AIPW estimator relies on conditional ignorability, positivity of the conditional treatment density $p(t|x)$, and continuous differentiability of the dose-response function $f$ and of $p$.  These are strong assumptions whose validity should be assessed on a case-by-case basis. However, a major benefit of IPW estimators is that they enable a suite of useful diagnostics, some of which we illustrate in Section~\ref{sec:empirical} \citep{austin2015moving}.

In our experience, a common way in which these assumptions can be violated is if the treatment is extremely sparse and/or skewed in regions of the covariate space.  Sparsity can be diagnosed by checking the distribution of estimated propensity scores.  Remedies for sparsity include trimming and/or grouping extreme treatment values \citep{petersen2012diagnosing}.  Indeed, in Appendix B of the supplementary material, we provide an MSE-based justification for grouping observations into relatively few bins (on the order of $N^{1/5}$).  In practice, researchers should assess the sensitivity of their results to this hyperparameter.  Plotting the treatment distribution and estimated dose-response curve, as we do in our empirical application, can also help reveal violations of positivity and continuity.

Another way to address treatment effect heterogeneity is to model Conditional Average Treatment Effects (CATEs) using pretreatment covariates \cite{Chernozhukov2018-fl, semenova2021debiased, wager2018estimation}.  For example, if one has pretreatment measures of treatment intensity (e.g., prior discount usage and/or other measures of price sensitivity), one can estimate how the treatment effect varies according to these covariates (e.g., the effect of a discount is larger for price-sensitive customers).

We note that this approach is \emph{complementary} to our own in the sense that one can also use the coarsened AIPW estimator to estimate CATEs in partitions of the covariates.  The conceptual difference is that we focus on treatment effect heterogeneity induced by nonlinear dose-response functions, whereas CATE methods focus on treatment effect heterogeneity explained by pretreatment covariates.

\section{Real-World Application}

\label{sec:empirical}

In this section, we demonstrate the empirical relevance of our theoretical analysis using real-world data from Netflix. The goal of this section is to show that the theoretical biases discussed above can (and, in our experience, often do) appear in real-world data.  Although we are unable to share the proprietary data analyzed in this section, we provide replication code at \url{https://github.com/winston-chou/rorr}.

In this application, we sought to understand how use of a feature, which we will call Feature A, affects future visits to Netflix.  To answer this question, we drew a random sample of 12.8m users and counted the number of times they used Feature A over a 28-day window.\footnote{Note that we only sampled users with at least one visit during this window.}  Next, we divided this number by each user's count of visits over the same period to define our continuous treatment, \texttt{Feature A Usage Rate}.  We defined our outcome as the count of each user's visits in the next 28-day window.  As covariates, we included the count of times each user used Feature A and the count of times each user visited the platform in the seven-, 14-, and 28-day windows preceding the treatment period.\footnote{We augmented these covariates with 25 additional covariates, including plays, engagement duration, and usage of other platform features in the 28 days preceding the treatment period.}

We divided our dataset into roughly equal-sized training, validation, and test datasets consisting of approximately 4.3 million users each.  To estimate the nuisance parameters in the PLM, we fit gradient boosted regression trees to the treatment and outcome variables observed in the training dataset, using the validation dataset to tune the number of boosting rounds.  Lastly, we regressed the outcome residuals on the treatment residuals in the test dataset to obtain the RORR treatment estimate.  This estimate is shown in the top row of Table~\ref{tbl:rorr-empirical}.

\begin{table}[h!]
\centering
\setlength{\tabcolsep}{10pt}
\begin{tabular}{lrrr}
\toprule
                   & RORR & Std. Err. & 95\% CI \\
\midrule
\textbf{Feature A Usage Rate} &     0.021  &        0.001     & (0.019, 0.023) \\
\midrule
                   & AIPW & Std. Err. & 95\% CI  \\
\midrule
\textbf{Feature A Usage Rate} & 0.788 & 0.011 & (0.766, 0.811) \\
\midrule
\multicolumn{4}{r}{$N = 4,324,101$} \\
\bottomrule
\end{tabular}
\medskip
\caption{RORR and AIPW Estimates of \texttt{Feature A Usage Rate} Treatment Effects.  Both the treatment and the outcome are divided by their respective standard deviations.}
\label{tbl:rorr-empirical}
\end{table}

In addition to the RORR estimate, Table~\ref{tbl:rorr-empirical} also reports the output of our coarsened AIPW estimator.  To fit this estimator, we coarsen the treatment into five bins and fit a multiclass classifier to these using gradient boosting.  We assign zero values (i.e., no usage of Feature A during the treatment period) to the first bin and then divide the remaining non-zero values into quartiles.  For each value of the coarsened treatment, we fit a separate outcome regression.

\begin{figure}[H]
    \centering
    \includegraphics[width=0.49\linewidth, keepaspectratio]{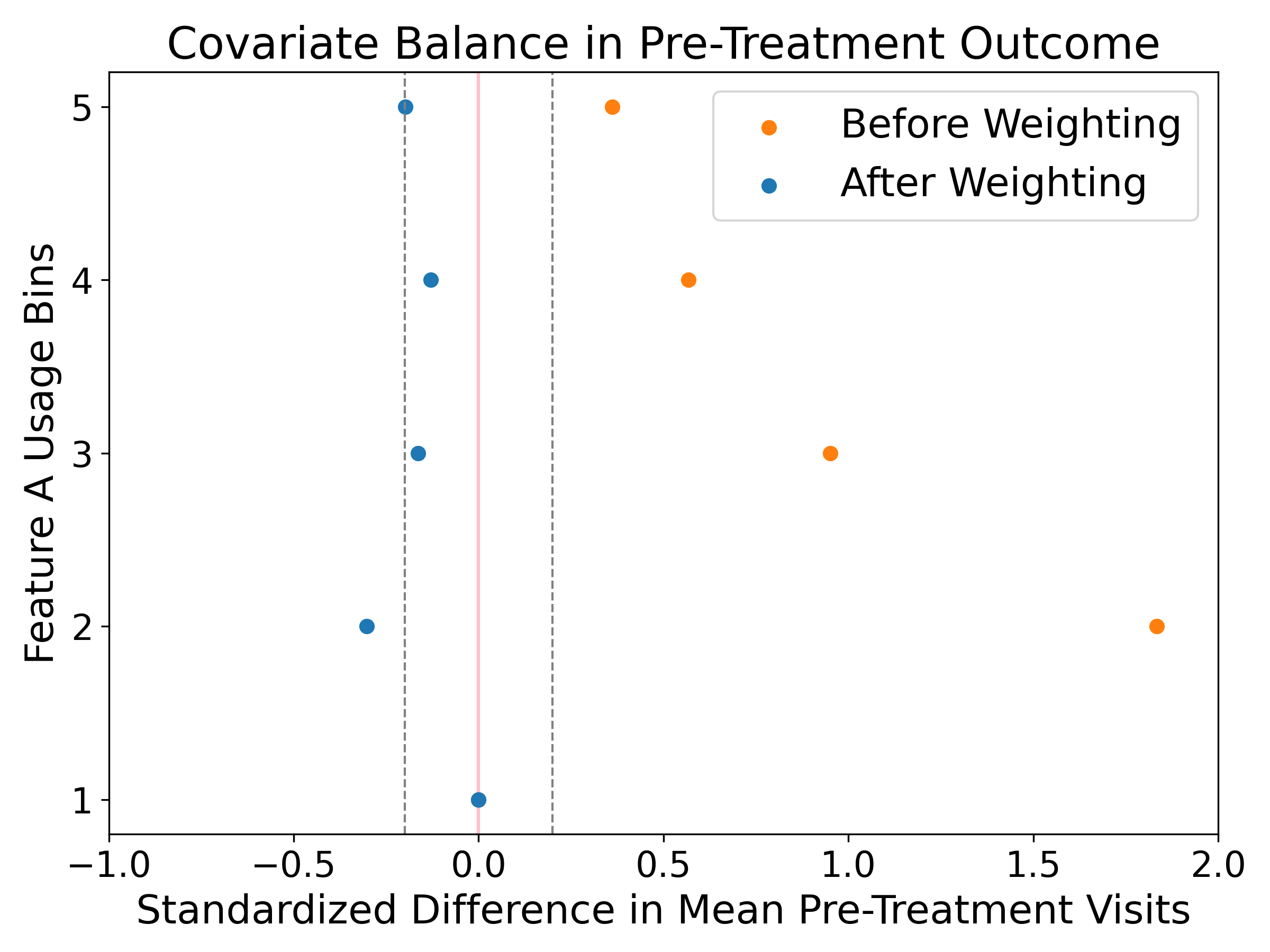}
    \includegraphics[width=0.49\linewidth, keepaspectratio]{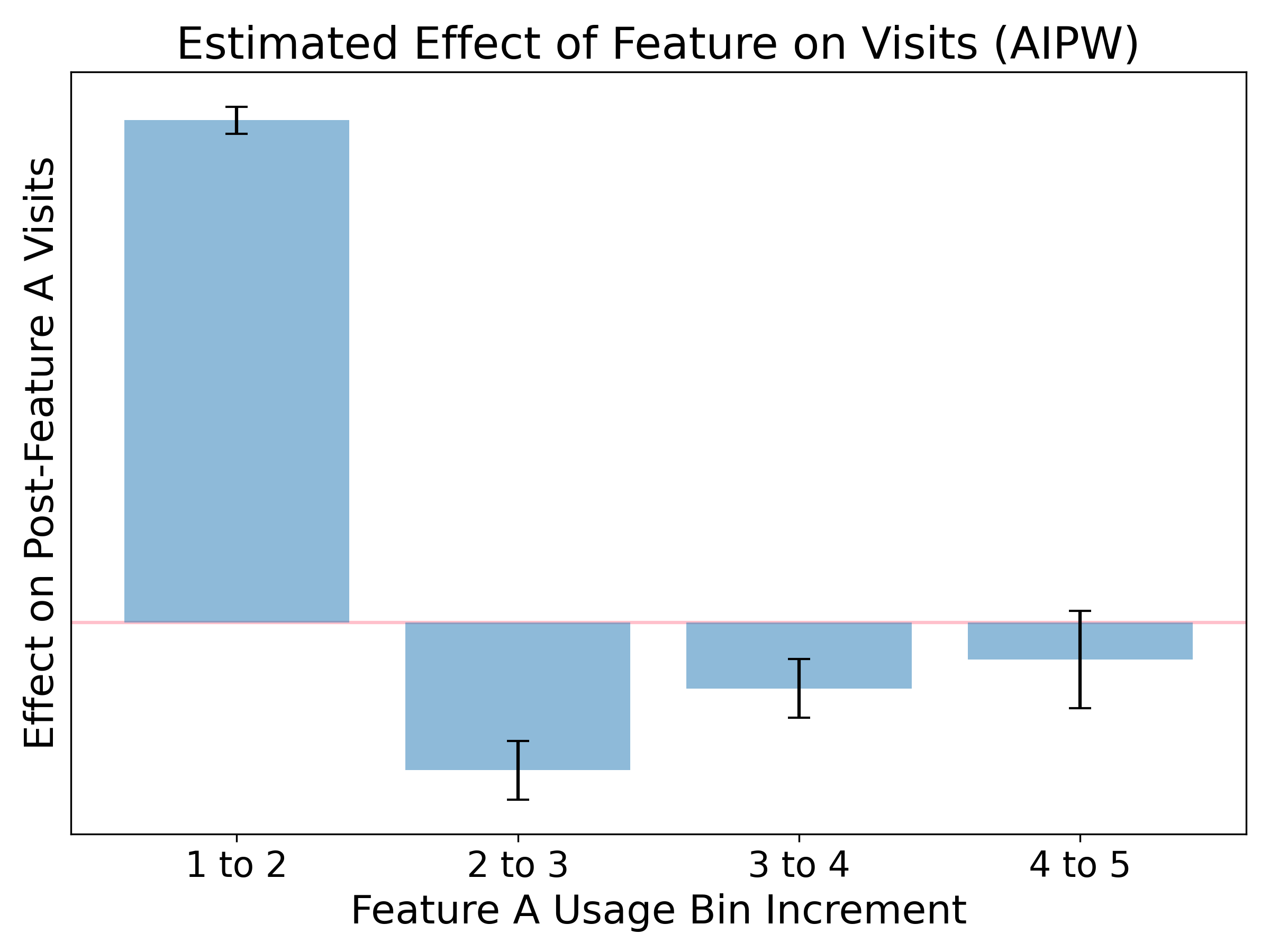}

    \includegraphics[width=0.49\linewidth, keepaspectratio]{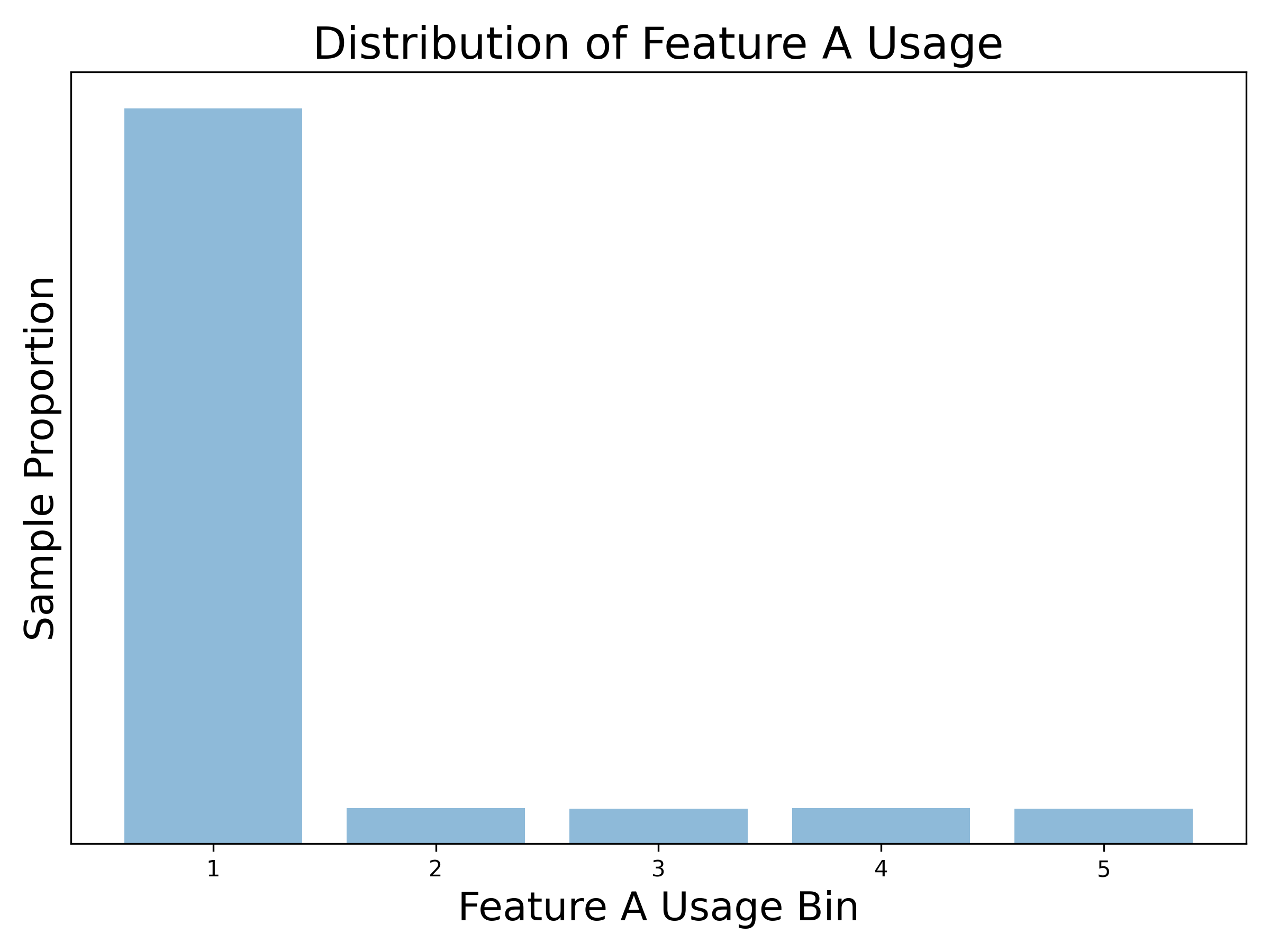}
    \includegraphics[width=0.49\linewidth, keepaspectratio]{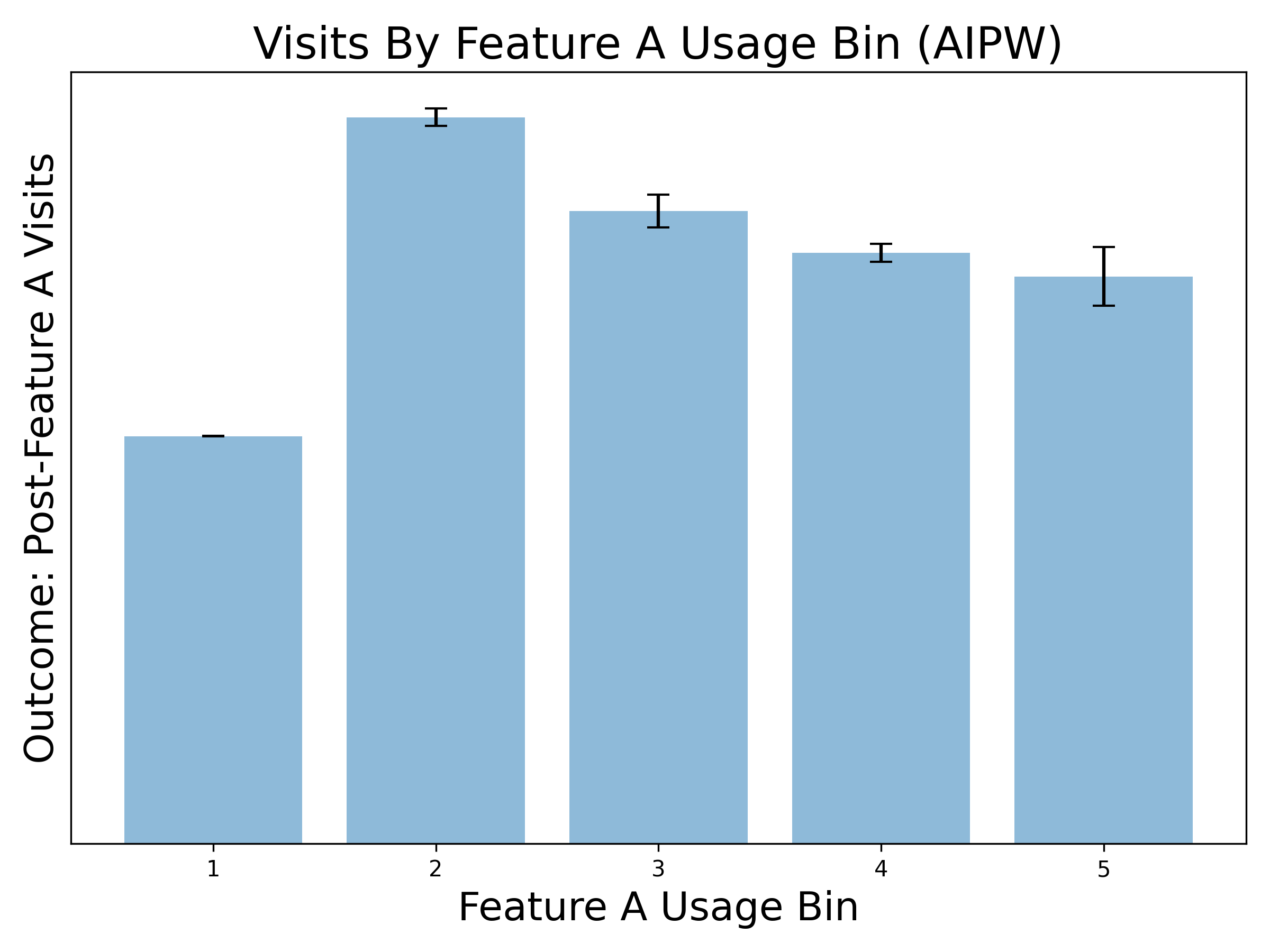}
    \caption{Estimating the Treatment Effect of \texttt{Feature A Usage Rate} on Netflix Visits with Coarsened AIPW. $y$-axis labels are hidden for confidentiality. Plots show, in clockwise order, covariate balance before and after weighting, incremental treatment effects, the dose-response curve, and the marginal treatment distribution. 
    }
    \label{fig:emp-results}
\end{figure}

As Table~\ref{tbl:rorr-empirical} shows, the coarsened AIPW estimate is substantially ($\sim$40x) larger than the RORR estimate.  To help explain this discrepancy, we plot the results of the AIPW estimator---and, crucially, the diagnostics it enables---in Figure~\ref{fig:emp-results}.

In the top-left panel of Figure~\ref{fig:emp-results}, we provide a standard IPW diagnostic plot of the difference in the standardized pretreatment value of the outcome in each bin and bin 1 before and after weighting.  As the panel shows, IPW significantly reduces pretreatment differences in the outcome variable, making the bins more comparable to each other and strengthening the plausibility of the unconfoundedness assumption.

In the top right panel, we plot the incremental effect of moving up one bin.  As the panel shows, AIPW estimates a large positive treatment effect of moving from the zero-usage bin (bin 1) to the next bin (bin 2), with negative incremental effects thereafter.  As the bottom-left panel shows, Feature A usage is highly zero-inflated, so bin 1 is the most representative bin.

Taken together, the diagnostics help to explain why the coarsened AIPW estimate is significantly larger than the RORR estimate:
\begin{itemize}
    \item The coarsened AIPW estimator explicitly weights the incremental treatment effects to be representative of the marginal treatment distribution, which is heavily zero-inflated.  Thus, it assigns the most weight to small values of the treatment, where the dose-response curve is steepest.
    \item Contrastively, RORR upweights units with more unpredictable values of the treatment, which tend to be larger.  This in turn overweights the downward-sloping portion of the dose-response curve.
\end{itemize}
The resulting attenuation bias of RORR is fully consistent with theoretical expectations.

The discrepancy in the estimated treatment effects is large and consequential for decision-making.  Although RORR indicates that Feature A has a negligible effect on visits, the AIPW results demonstrate that, for the vast majority of users, increasing Feature A usage would have a much larger effect.  Moreover, as the incremental treatment effects panel shows, a negligible positive effect is not representative of \emph{any} bin-level treatment effect in the data.

\section{Conclusion}

Double Machine Learning estimators are becoming increasingly popular in both academic and commercial applications of observational causal inference.  Focusing on the residuals-on-residuals regression (RORR), we analyze the bias of RORR when treatment effects are heterogeneous.  For binary treatments, RORR estimates a weighted average of treatment effects that is skewed towards units for whom the conditional probability of treatment is closest to 0.5.  For non-binary treatments, RORR converges to a weighted average of causal derivatives, with the added complication that these derivatives are evaluated on a ``pseudo-treatment'' distribution that is not the treatment distribution found in the data.

As our empirical application shows, the bias of RORR relative to the average treatment effect can have significant consequences for decision-making.  As an alternative, we propose a doubly-robust Augmented Inverse Propensity Weighting estimator and show theoretically and empirically that this yields more representative estimates of causal effects.

\begin{credits}
\subsubsection{\ackname} For valuable suggestions, we thank the anonymous reviewers, Peter Hull, Yi Zhang, and participants in the Workshop on Causal Inference and Machine Learning in Practice at KDD '25 in Toronto, CA.  For building and maintaining the observational causal inference platform that utilizes the methods described in this paper, we thank Adrien Alexandre, Colin Gray, and Dan Zylberglejd.  The Appendix for this paper can be found at \url{https://github.com/winston-chou/rorr}.

\subsubsection{\discintname}
The authors have no competing interests to declare that are
relevant to the content of this article.
\end{credits}
%
%
%
\bibliographystyle{splncs04}
\bibliography{bib}

\end{document}